\documentclass[journal]{IEEEtran}

\usepackage{cite}
\usepackage{amsmath,amssymb,amsfonts}
\usepackage{algorithmic}
\usepackage{graphicx}
\usepackage{textcomp}

\usepackage{epstopdf}
\usepackage[utf8]{inputenc}
\usepackage[caption=false,font=footnotesize]{subfig}
\graphicspath{{images/pdf/}}
\usepackage{array}
\usepackage{upgreek}
\usepackage{color,soul}
\usepackage[version=3]{mhchem}
\usepackage{arydshln}
\usepackage{comment}
\usepackage{bm}
\usepackage{hhline}
\usepackage{float}
\usepackage{placeins}
\usepackage{afterpage}

\begin{document}

\title{Parameterizing the Angular Distribution of Emission: A Model for TOF-PET Low-Count Reconstruction}

\author{Maxime~Toussaint,
		Francis Loignon-Houle,
		Roger~Lecomte,
		and~Jean-Pierre~Dussault%
		\vspace{-2mm}
\thanks{This work was supported by Discovery grants from the Natural Sciences and Engineering Research  Council of Canada (NSERC), Mitacs and the MEDTEQ Consortium of the \textit{Ministère de l'économie, de la science et de l'innovation} (MÉSI) of the Government of Québec. 
This research was possible in part by support provided by \textit{Calcul Québec} and Compute Canada.
The Sherbrooke Molecular Imaging Center is a member of the FRQS-funded \textit{Centre de Recherche du CHUS} (CRCHUS).
}
\thanks{M. Toussaint and J.-P. Dussault are with the Department of Computer Science, Université de Sherbrooke, Sherbrooke, QC, Canada.}
\thanks{F. Loignon-Houle and R. Lecomte is with the Sherbrooke Molecular Imaging Center of CRCHUS and Department of Nuclear Medicine and Radiobiology, Université de Sherbrooke, Sherbrooke, QC, Canada.}
}

\maketitle

\begin{abstract}
	Low-count reconstruction remains a challenge for Positron Emission Tomography (PET) even with the recent progress in time-of-flight (TOF) resolution.
	In this context, the bias between the acquired histogram, consisting of low values or zeros, and the expected histogram, obtained from the forward projector, is propagated to the image, resulting in a biased reconstruction.
	This situation could be exacerbated with finer resolution of the TOF information, which further sparsifies the acquired histogram.
	We propose a new approach to circumvent this limitation of the classical reconstruction model.
	It consists of extending the description of the reconstruction scheme to also explicitly include the projection domain by Parameterizing the Angular Distribution of Emission (PADE).
	This parametrization has greater degrees of freedom than the log-likelihood model, which can not be harnessed in classical circumstances.
	We hypothesize that with ultra-fast TOF this new approach would not only be viable for low-count reconstruction, but also more adequate than the classical reconstruction model.
	As a proof of concept, an implementation of this approach is compared to the log-likelihood model by investigating two-dimensional simulations of a hot spots phantom.
	The proposed model achieves similar contrast recovery coefficients as MLEM except for the smallest structures where the low-count nature of the simulations makes it difficult to draw conclusions.  
	However, this new model seems to converge toward less noisy solutions than MLEM.
	These results suggest that the PADE approach has potential for low-count reconstruction with ultra-fast TOF.
\end{abstract}


\bstctlcite{IEEEexample:BSTcontrol}

\section{Introduction}
	\IEEEPARstart{T}{he} gain in image quality provided by time-of-flight (TOF) in positron emission tomography (PET) is directly linked to coincidence time resolution (CTR)~\cite{conti2011focus,SURTI201612,vandenberghe2016recent}. 
	Recently, progress achieved in PET detectors have made it possible to build a clinical scanner with a CTR approaching 200~ps~\cite{van2019performance}.
	Further advances can be expected since a CTR of 58~ps was recently achieved, albeit in a benchmark test using short crystals of LSO:Ce:0.4\%Ca~\cite{gundacker2020experimental}.  
	Moreover, the roadmap towards achieving a CTR of 10~ps has been laid out and technological solutions have been discussed extensively~\cite{lecoq2020roadmap}.
	The challenge is huge and many pitfalls remain~\cite{schaart2020achieving}.
	Therefore, a better understanding of the benefits provided by ultra-fast TOF is of interest.
	For example, it is well known that TOF improves the signal-to-noise ratio in PET images~\cite{budinger1983Time,conti2011focus,conti2013estimating}.
	In addition, it was recently demonstrated that ultra-fast TOF resolution could provide a gain in spatial resolution by mitigating the blur induced by the detectors size~\cite{toussaint2020improvement}.
	Our goal is to investigate how ultra-fast TOF can be exploited in the PET reconstruction scheme to yield further gain in image quality and robustness.
	
	Low-count acquisition remains a challenge for PET reconstruction~\cite{6878472,lim2018pet}.
	When the correction for random events is applied, the non-negativity constraints of the MLEM reconstruction scheme induces a bias in the evaluation of the projection space correction factors that is propagated in the reconstructed image.
	A solution proposed in~\cite{6878472} consists of replacing the usual Poisson distribution by a Gaussian distribution for projections with a low number of counts.
	The authors of~\cite{lim2018pet} propose a new reconstruction scheme in which the non-negativity constraint is shifted from the image space to the projection space. 
	In both cases, negative values in the image space are permitted to circumvent the bias induced by the estimation of random coincidences included in the projections with low number of counts.
	While one could expect the random coincidences to dwindle with better TOF resolution, the coincidence time window will always be lower bounded by the subject size.
	Another indication of the limitation of the classical PET model for low-count acquisition is highlighted in~\cite{westerwoudt2014advantages}, where it is shown that the TOF filtered back projection could outperform MLEM in terms of signal-to-noise ratio.
	
	Since low spatial frequencies converge faster than higher frequencies with the MLEM algorithm, its convergence rate depends on the structures size.
	Over-iteration with the MLEM algorithm also results in noisy images~\cite{qi2006iterative}.
	Noise can thus contaminate larger structures before smaller ones can reach their optimal contrast.
	This behavior compels users to arbitrarily terminate the reconstruction process early, based on the structures of interest.
	This could be exacerbated with ultra-fast TOF since convergence rate increases with better TOF resolution~\cite{SURTI201612}. 
	A regularization term can be included in the reconstruction model to mitigate noise~\cite{qi2006iterative}.
	However, the optimal solution of a low-count reconstruction using a fine image discretization includes high frequency structures, which can impede the efficiency of a regularization scheme.
	
	In PET models, the physical property of uniform emission distribution from a point source in the 3D sphere is encoded in the system matrix.
	Therefore, the forward model promotes a predetermined ratio of the counts in a voxel to all its projections, irrespective of the observed data. 
	We propose to extend the parametrization of the reconstruction scheme to also explicitly include the projection domain in order to circumvent the bias induced by statistical noise in low-count reconstruction.
	The resulting number of variables is increased by an order of magnitude relative to the classical approach. 
	Nevertheless, we hypothesize that with ultra-fast TOF, the proposed approach can be well-defined and, thus, provide a better description of the underlying physical processes for the reconstruction of low-count acquisition. 
	
	An implementation of this new approach for low-count acquisitions was investigated as a proof of concept.
	The image quality achieved by this model was studied with a simulated 2D hot spots phantom.
	The MLEM algorithm was used as a baseline.  
	Overall, the results of this study support the hypothesis that the proposed approach is of interest for low-count reconstruction with ultra-fast TOF.

\section{Parameterizing of the Angular Distribution of Emission (PADE) Model}
\label{sec:model}
	In this section, we propose an implementation of this new approach to PET reconstruction.
	Correction factors associated to attenuation, detector efficiency, randoms and scatters are omitted for simplicity.
	The Python convention was used when a subset of a tensor is taken, e.g., $P_{:,i}^t$ is equivalent to $\{P_{j,i}^t| j \in [1, J]\}$.
	Also, the term pixel was used interchangeably for voxel since a part of the model is described for 2D reconstruction. 
	Let $y_j^t$ be the number of coincidences observed in projection $j$ at the TOF bin $t$ and $P$ the TOF system matrix of a scanner.
	The likelihood model consists of solving $y_j^t\sim\textrm{Poisson}( \sum_i P_{j,i}^t \lambda_i), \forall j, t$ with $\lambda_i$ being the number of coincidences originating from the pixel $i$. 
	The expected contribution of pixel $i$ to projection $j$ and TOF bin $t$ for a given estimate $\lambda$ is $P_{j,i}^t \lambda_i$.
	To fully exploit ultra-fast TOF resolution, the TOF information needs to be finely discretized.
	This results in $y$ being sparse and low counts, which makes its approximation with $P_{j,i}^t \lambda_i$ highly prone to bias.

	We propose to extend the parametrization of the reconstruction model to include the projection space. 
	Let $\phi_{j, i}$ define the number of coincidences emitted from pixel $i$ and observed in projection $j$.
	In that case, $\lambda_{i} = \sum_{j}{\phi_{j, i}}$ and the expected contribution of $\phi_{j,i}$ to $y_j^t$ is $Q_{j,i}^{t} \phi_{j, i}$, where $Q_{j,i}^{t}$ is the probability of a coincidence originating from pixel $i$ and observed in projection $j$ to be observed in TOF bin $t$.
	Consequently, $Q_{j,i}^{t} R_{j,i} \approx P_{j,i}^t $ where $R$ is the TOF-less system matrix of the scanner.
	Introducing this parameterization in the log-likelihood model results in
	\begin{equation}\label{eq:likelihoodExt}
		\mathcal{L}(\phi) = \sum_{jit} Q_{j, i}^{t} \phi_{j, i} - \sum_{jt} y_j^t \ln\left(\sum_i Q_{j, i}^t \phi_{j, i} \right)
	\end{equation}
	which is the data fit term for the Parameterizing of the Angular Distribution of Emission (PADE) model, explicitly introducing the emission angular distribution in projection space.
	$\phi$ can be defined as a sparse matrix along the projection dimension since a given pixel intersects only a very small subset of the projections spanning the entire image domain.
	Thus, the proposed extension does not multiply the number of variables by the total number of projections defined in the scanner but does provide a significant increase in degrees of freedom.
	Also, one could note that the proposed parametrization corresponds to the latent variables used in the MLEM formalism when applied to TOF-less PET reconstruction~\cite{Lange1984}.

	A core property of the PET system model is missing in~\eqref{eq:likelihoodExt}: the isotropic nature of its emission that we refer to as the uniform distribution of emission (UDE) property.
	This is usually enforced by the system matrix as $ \phi_{j,i} = R_{j,i} \lambda_i, \forall j,i$.
	Let $\mathcal{V}_i\left( \phi \right)$ be a function that numerically evaluates how closely the pixel $i$ follows the UDE property.
	In this study, the UDE property is enforced as a penalization term defined as
	\begin{equation}\label{eq:geoPenal}
		\mathcal{U}(\phi) = \sum_{i} \omega_i \mathcal{V}_i\left( \phi \right),
	\end{equation}
	where $\omega_i$ is the weight applied to pixel $i$.
	
	\begin{figure}
		\centering
		\subfloat[Partition of a point source emission angles over projections]{%
			\label{fig:emissionPartInScanner}\includegraphics[width=0.45\linewidth]{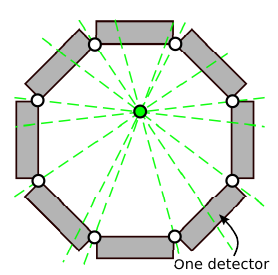}}
		\quad
		\subfloat[Cones of emission and two examples of $j_i^{\textrm{ref}}$ for the point source]{%
			\label{fig:coneOfEmission}\includegraphics[width=0.45\linewidth]{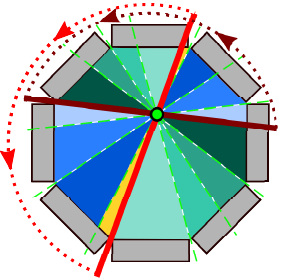}}\\
		\subfloat[Polar partition of the emission angles relative to the dark red $j_i^{\textrm{ref}}$]{%
			\label{fig:vizCircBothPartition}\includegraphics[width=0.45\linewidth]{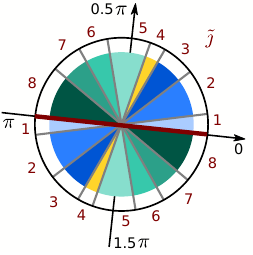}}
		\quad
		\subfloat[Visualization of three counts in the previous partition]{%
			\label{fig:vizCircBothPartitionExample}\includegraphics[width=0.45\linewidth]{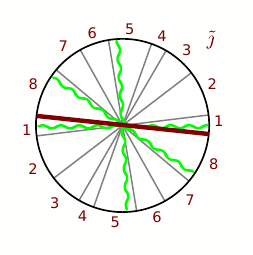}}
		\caption[Representation of the cyclic nature of PET emission and the impact of $j_i^{\textrm{ref}}$ for an octogonal-shaped scanner with one detector per panel]{%
			Representation of the cyclic nature of PET emission and the impact of $j_i^{\textrm{ref}}$ for an octogonal-shaped scanner with one detector per panel.
			The white circles in \protect\subref{fig:emissionPartInScanner} were added to highlight that the cones of emission in \protect\subref{fig:coneOfEmission} are obtained by extending the lines that connect the detectors edges to the point source, represented by the dashed green lines.
			The resulting eight cones of emission were colored differently for easier visualization.
			The dark red and light red lines indicate two possible $j_i^{\textrm{ref}}$ for ordering the valid projections.
			\protect\subref{fig:vizCircBothPartition} shows the partition of $[0, 2 \pi)$ from the cones of emission of the point source, when ordered using the dark red $j_i^{\textrm{ref}}$.
			The partition of $[0, \pi)$ is sufficient to define the cones of emission due to the anti-parallel nature of PET emission. 
			\protect\subref{fig:vizCircBothPartitionExample} shows an example of three emission angles that correspond to one count for $\tilde{\jmath} = \{1, 5, 8\}$ and zero everywhere else for the dark red $j_i^{\textrm{ref}}$.
			For the light red $j_i^{\textrm{ref}}$, the counts would be in $\tilde{\jmath} = \{1, 4, 5\}$.
		}
		\label{fig:visuUde_geo}
	\end{figure}  	
	
	The implementation of $\mathcal{V}\left( \right)$ was inspired by section 4.4.1 of~\cite{mardia2009directional} where it is pointed out that the uniform distribution is the only circular distribution that is invariant under rotation. 
	It means that the stability over rotations of a parameter evaluated on a given dataset can be used to determine if the dataset follows a circular uniform distribution.
	The word stable, compared to invariant, is important: it is the expected value of the parameter that is invariant, not its estimation for a given sample.
	We decided to use a weighted sum, that we refer to as momentum, as a numerical characterization of the pixel projection-wise distribution.
	The momentum, defined in the next paragraphs, characterizes the central tendency of a dataset relative to an angle of reference.
	For a circular uniform distribution, the expected momentum is the middle of its angular domain.
	It was chosen rather than the mean since the latter requires a division by the number of elements in the distribution, $\sum_j \phi_{j,i} $, making $\mathcal{V}\left( \right)$ more complex and the resulting model harder to solve.
	However, multiple types of distributions can share a specific central tendency for a given angle of reference: it is by using multiple angles of reference that the momentum will be able to characterize whether a dataset follows a circular uniform distribution or not.
	
\begin{figure}
		\centering
		\subfloat[Computation of $\mathcal{W}_i\left( \phi, j_i^{\textrm{ref}} \right)$ for the dark red $j_i^{\textrm{ref}}$]{%
			\label{fig:udePenalViz_exA}\includegraphics[width=0.95\columnwidth]{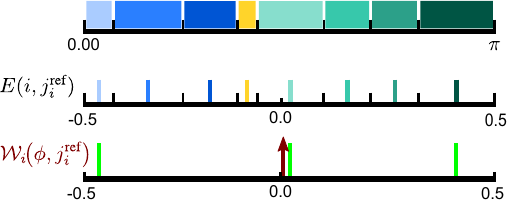}}\\
		\subfloat[Computation of $\mathcal{W}_i\left( \phi, j_i^{\textrm{ref}} \right)$ for the light red $j_i^{\textrm{ref}}$]{%
			\label{fig:udePenalViz_exB}\includegraphics[width=0.95\columnwidth]{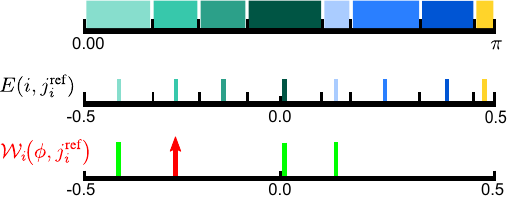}}
		\caption[Representation of the process behind~\eqref{eq:geoPenalMetricInnerLoop} using the example in Fig.~\ref{fig:visuUde_geo}]{%
			Representation of the process behind~\eqref{eq:geoPenalMetricInnerLoop} using the example in Fig.~\ref{fig:visuUde_geo}.
			\protect\subref{fig:udePenalViz_exA} represents the case of the dark red $j_i^{\textrm{ref}}$ in three steps.
			The first line shows the same partition as Fig.~\ref{fig:vizCircBothPartition} but limited to $[0, \pi)$.
			The second line shows the resulting $E(i, j_i^{\textrm{ref}})$, represented by lines colored to their corresponding projection. 
			The third line shows the bins where $\phi_{j,i} = 1$ as green lines and the dark red arrow shows the result of $\mathcal{W}_i\left( \phi, j_i^{\textrm{ref}} \right)$.
			\protect\subref{fig:udePenalViz_exB} is the same as previous except with the light red $j_i^{\textrm{ref}}$.
			Note that the position of the resulting arrow differs between the two choices of $j_i^{\textrm{ref}}$ and that the position for the light red $j_i^{\textrm{ref}}$ illustrates the counter-intuitive aspect of using the momentum instead of the mean in~\eqref{eq:geoPenalMetricInnerLoop}. 
		}
		\label{fig:visuUde_num}
	\end{figure}

	The function $\mathcal{V}\left( \right)$ implemented in this study needs the position of the projections in the pixel emission space, i.e., in the circle for 2D reconstructions, to characterize numerically the pixel distribution.
	This is unusual in PET reconstruction and it requires extra care in its description.
	Therefore, a visual representation of the key steps behind the computation of $\mathcal{V}\left( \right)$ are provided in Fig.~\ref{fig:visuUde_geo} and Fig.~\ref{fig:visuUde_num}.
	The goals of these steps are to define an ordering for the projections, to circumvent the discrete sampling of the projection domain and to modify the emission angular space so that it is centered at zero.
	
	$\mathcal{V}\left( \right)$ is implemented as follow.
	A partition of the emission angular space is obtained by using the lines that connect the detectors extremities to a pixel, as illustrated in Fig.~\ref{fig:emissionPartInScanner}.
	We refer to a part of that partition as a cone of emission, which is kind-of the dual of the TOF-less tube of response ($R_{j, :}$).
	The cones of emission of the pixel $i$ ($R_{:, i}$) in Fig.~\ref{fig:emissionPartInScanner} are highlighted in Fig.~\ref{fig:coneOfEmission}.
	Let $j_i^{\textrm{ref}}$ be a projection of reference for pixel $i$ from which valid projections, i.e. $\{j | R_{j,i} \neq 0.0\}$, can be ordered.
	The choice of $j_i^{\textrm{ref}}$ emulates a rotation over the emission angular space, as shown with the two examples of $j_i^{\textrm{ref}}$ given in Fig.~\ref{fig:coneOfEmission}, highlighted with the light red and the dark red lines.
	The resulting polar partition of the emission angular space for the dark red $j_i^{\textrm{ref}}$ is shown in Fig.~\ref{fig:vizCircBothPartition}.
	An example of the distribution obtained for three emissions originating from the pixel is illustrated in Fig.~\ref{fig:vizCircBothPartitionExample} for the polar partition of Fig.~\ref{fig:vizCircBothPartition}.
	The first line of Fig.~\ref{fig:udePenalViz_exA} and~\ref{fig:udePenalViz_exB} show the partition of the emission angular space obtained from the example in Fig.~\ref{fig:visuUde_geo}, respectively for the dark red and light red $j_i^{\textrm{ref}}$.
	Its discretization was defined as the middle position of the partition resulting from the choice of $j_i^{\textrm{ref}}$.
	Let $D(i, j_i^{\textrm{ref}})$ be a vector that holds that discretization and let $E(i, j_i^{\textrm{ref}}) = D(i, j_i^{\textrm{ref}}) / \pi - 0.5$ be the discretization employed by $\mathcal{V}\left( \right)$, illustrated in the second line of Fig.~\ref{fig:udePenalViz_exA} and~\ref{fig:udePenalViz_exB}.
	The subscript $\tilde{\jmath}$ will be used for $E(i, j_i^{\textrm{ref}})$ to specify that the order of the projections differs from $j$ and that only a subset of the projections are defined for a given pixel.
	Let $\mathcal{C}_i(\tilde{\jmath}, j_i^{\textrm{ref}})$ be a function that provides the projection index of the $\tilde{\jmath}$-th projection relative to $j_i^{\textrm{ref}}$ for pixel $i$.
	Thus, the momentum, i.e. a custom weighted sum, of the distribution $\phi_{:, i}$, which is not a mean, is 
	\begin{equation}\label{eq:geoPenalMetricInnerLoop}
		\mathcal{W}_i\left( \phi, j_i^{\textrm{ref}} \right) = \sum_{\tilde{\jmath}} \left( [E(i, j_i^{\textrm{ref}})]_{\tilde{\jmath}} ~ \phi_{l,i} \right)
	\end{equation}	
	with $j = \mathcal{C}_i(\tilde{\jmath}, j_i^{\textrm{ref}})$. 
	This is evaluated over several projections of reference to incorporate its variation over rotations.
	Thus, the UDE penalization term employed is 
	\begin{equation}\label{eq:geoPenalMetric}
		\mathcal{V}_i\left( \phi \right) = B^{-1} \sum_{j_i^{\textrm{ref}}} \left( \mathcal{W}_i\left( \phi, j_i^{\textrm{ref}} \right) \right)^2,
	\end{equation}
	where $B$ is the number of $j_i^{\textrm{ref}}$ considered.
	If a distribution follows exactly the UDE property, the momentum will be zero for all projections of reference.
	Fig.~\ref{fig:visuUde_num} shows a case where~\eqref{eq:geoPenalMetric} vary with the choice of $j_i^{\textrm{ref}}$.
	The value being almost zero for the dark red $j_i^{\textrm{ref}}$ (Fig.~\ref{fig:udePenalViz_exA}) contrary to the value obtained with the light red $j_i^{\textrm{ref}}$ (Fig.~\ref{fig:udePenalViz_exB}). 	
	
	One can expect that the UDE penalization term will be more complex for 3D reconstructions since emission angles can no longer be represented over a circle. 
	However, most scanners are shaped as a stack of rings and thus the sampled emission space of a voxel will be a stack of rings along one of the axis of the 3D sphere.
	The resulting sampling schemes can be mapped to a 2D uniform space if solid angles are taken into accounts.
	Then, the methodology defined above should be adequate as long as considerations of a plane domain over a linear domain are taken into account (e.g. the momentum would be in 2D).
	Since most PET sampling schemes exclude two large spherical caps, we expect that the cyclic nature of the UDE might only be usable for the polar angles, with the zenith direction being the axial axis of the PET camera. 
	
	A convenient property of the MLEM algorithm is the stability of the expected number of coincidences over all iterations~\cite{qi2006iterative}.
	Let $\lambda^k$ be the image obtained after $k$ iterations of MLEM, then $\sum_{t,j,i} P_{j,i}^{t}\lambda_i^k = \sum_{t,j} y_j^t, \forall k > 1$.
	However, this property needs to be incorporated in the reconstruction model for general solvers.
	It can be enforced as a constraint, with a penalization term or within the solver update scheme.
	We used a penalization term defined as
	\begin{equation}\label{eq:countsPenal}
		\mathcal{H}(\phi) = \left( \sum_{t,j,i} Q_{j,i}^t \phi_{j,i}  - \sum_{t,j} y_j^t \right)^2.
	\end{equation}

	Thus, the general form of the PADE model that was explored was
	\begin{equation}\label{eq:padeModel}
		\begin{aligned}
			&\min_{\phi}       & \qquad & \mathcal{L}\left(\phi\right) + \gamma_1 \mathcal{U}\left(\phi\right) + \gamma_2 \mathcal{H}\left(\phi\right)  \\
			&\text{subject to} &      & \phi \geq 0.0,
		\end{aligned}
	\end{equation}
	where $\gamma_1$ is the weight for the UDE penalization term and $\gamma_2$ the weight for the expected number of coincidences penalization term, with $\gamma_1, \gamma_2 \geq 0.0$.
	Compared to the classical PET log-likelihood model, the model described in~\eqref{eq:padeModel} relaxes the interpretation of the UDE property by replacing the relation $\phi_{j,i} = R_{j,i} \lambda_i$ by a penalization term represented by $\mathcal{U}\left(\phi\right)$. 
	This new model requires more care to solve than the classical PET log-likelihood model.
	For example, the MLEM inherently imposes non-negativity constraints on $\lambda$.
	This is not the case for all general solvers and it will thus need to be taken into account in the choice of the solver.
	Another example is the calibration of $\gamma_1$ and $\gamma_2$: their values affect the quality of the solutions and can be data-dependent.


\section{ Simulation setup }
\label{sec:meth}
	The goal of this study was to show that the PADE approach, which relaxes the interpretation of the uniform distribution of emission (UDE) property, is a legitimate candidate for low-count reconstruction. 
	The methodology was built to compare the image quality achieved by the proposed PADE implementation to the performance of the MLEM algorithm on low-count acquisitions without including further considerations.

	\subsection{ Data simulation }
	\label{subsec:dataSimul}
		The version 8.0 of Geant4 Application for Tomographic Emission (GATE)~\cite{jan2004gate} was used to simulate data acquisition from a fictive PET ring scanner.
		The sources were defined as back-to-back which means that positron range and annihilation photon acolinearity were not simulated.
		The emission direction of the sources was limited to the 2D plane of the scanner.
		Only photoelectric processes were enabled for the annihilation photons hence the datasets does not have scatter coincidences.
		Also, the emission rate of the sources was chosen such that random coincidences would be negligible.
	
		The 2D ring camera was shaped as a regular polygon with 40 sides and an inner diameter of $\sim$80~cm.
		Each panel was 64~mm in size and consisted of 8 detectors of 8~mm in width. 
		The scanner thus had 320 detectors, implying that each image pixel was intersected by around 320 tubes of response (i.e. valid projections).	
		The detector length was fixed to 0.1~mm, making the blur induced by depth of interaction negligible. 
		The detectors were 4~mm wide in the axial axis and it was assumed that they had a direct readout without light-sharing decoding.
		The CTR was fixed to $\sim$13~ps at full width at half maximum (FWHM), resulting in a TOF spatial resolution of 2~mm along projections. 

		A custom hot spots phantom was created for this study.  
		Its main body was a cylinder of 136~mm in diameter and 4~mm in height.
		The hot spots diameters were 3.2, 4.8, 6.5, 7.9, 9.5, 11.1~mm.
		The activity in the phantom was defined such that the contrast between the spots and the background would be of four.
		The simulation was repeated ten times, each resulting in a dataset of around 80,000 coincidences.

	\subsection{ Image reconstruction }
	\label{subsec:methRecon}
		The image domain was a $16\times16$~cm$^2$ plane discretized uniformly in $128 \times 128$ pixels, resulting in an in-plane pixel size of $1.25\times1.25$~mm$^2$.
		A total of 8,544 projections intersected the image domain.
		The TOF information was discretized uniformly in 128 bins resulting in spatial bins of 1.82~mm in width along the tubes of response.
		Thus, the histograms consisted of 1,093,632 bins and most of them had zero coincidences.
		
		The TOF, $P$, and TOF-less, $R$, system matrices and the pure-TOF matrix, $Q$, were precomputed.
		$R_{j,i}$ was approximated as the geometric probability of a point source centered in the voxel $i$ to emit in the tube of response of projection $j$.
		The TOF response function of a projection was assumed invariant across its tube of response and modeled as a 1D Gaussian of 2~mm FWHM along its tube of response.
		$Q_{j,i}^t$ was approximated as the result at voxel $i$ of the convolution of the TOF response function with the rectangular function associated to the TOF bin $t$. 
		$P^t_{j,i}$ was defined as $Q^t_{j,i} R_{j,i}, \forall t,j,i$.
		For all three matrices, the image domain was oversampled three times in both dimensions of the plane (i.e. nine samples per pixel) and the mean was taken.
		
		The L-BFGS-B~\cite{2020SciPy-NMeth,zhu1997algorithm} algorithm with non-negativity constraints was employed to solve the PADE model.
		The parameter of limited memory was set to 10. 
		A reduction of the number of variables was applied following the idea that variables having a null probability of being associated with any coincidences are irrelevant.
		Thus, the constraints were modified from $\phi_{j,i} \geq 0, \forall j,i$ to 
		\begin{equation}\label{eq:constraints}
			\begin{cases}
			    \phi_{j,i} = 0.0 & \text{if } \sum_{t} y_j^t P^t_{j,i} = 0.0 \\
			    \phi_{j,i} \geq 0.0              & \text{otherwise.}
			\end{cases}
		\end{equation}
		In a ultra-fast TOF and low counts setting, a significant portion of the variables can be deactivated with this approach ($\approx$70\% in this study).
		The pixel weights for the UDE penalization (i.e. $\omega$ of~\eqref{eq:geoPenal}) were defined in two steps.
		First, it would be detrimental to penalize pixels that have a low value since the UDE property can not be evaluated accurately in those cases.
		For this proof of concept, we applied a hard threshold for pixels lower than 9.0 since it was the highest value found in the body of the phantom for the simulation groundtruth (i.e. the true number of emissions per pixel). 
		Second, the strength of the penalty applied on a pixel should be positively correlated with its value.
		We chose to set the weights at each iteration as the value of the pixels of the previous iteration which was inspired from the approach used in the OSL-MLEM algorithm~\cite{green1990bayesian}.
		Thus, the pixel-dependent weights were defined as
		\begin{equation}\label{eq:omegaWeight}
			\omega_i^k = 
			\begin{cases}
			    0.0 & \text{if } \sum_j \phi_{j,i}^0 < 9.0 \\
			    \sum_j \phi_{j,i}^{k-1}             & \text{otherwise.}
			\end{cases}
		\end{equation}		
		Our initial tests had shown that the proposed model did not perform well when it was initialized with an uniform image.
		This might have been due to the update scheme of $\omega$ which results in the objective function being drastically modified at each iteration.
		However, we observed that the MLEM model could produce an adequate initialization for the proposed model, if stopped early.
		Therefore, a low iteration MLEM reconstruction, $\lambda^k$ with $k$ being the number of iterations, was employed to build an initial estimate.
		Since $\lambda^k$ is only defined in the image domain, the initial estimate, $\phi^0$, was generated using the TOF-less system matrix.
		A modified version of the $R$ matrix was employed in order to satisfy~\eqref{eq:constraints}.
		Let $\widetilde{R}_{j,i}$ be equal to $R_{j,i}$ if $\sum_t y_j^t P^t_{j,i} \neq 0.0$ and 0.0 otherwise.
		The matrix was then normalized such that $\sum_j \widetilde{R}_{j,i} = \sum_j R_{j,i}, \forall i$.
		Thus, the PADE model was initialized with 
		\begin{equation}\label{eq:initialization}
			\phi_{j,i}^0 = \widetilde{R}_{j,i} \lambda_i^k, \forall j,i.
		\end{equation}
		The iteration that had the minimum mean squared error with the simulation groundtruth was chosen, which was achieved at the 6$^{\textrm{th}}$ iteration when looking at the mean over all repetitions.
		An advantage of using an image obtained from MLEM for initialization is that it has the correct number of expected coincidences.
		Since the number of valid projections was around 320 for all pixels, the computation of~\eqref{eq:geoPenalMetric} over all possible $j_i^{\textrm{ref}}$ was computationally expensive. 
		In the exploration of the proposed model (not shown), only subsets of 10, 20 and 30 $j_i^{\textrm{ref}}$ were considered due to the computational requirement for the calibration of the weights $(\gamma_1, \gamma_2)$ (described in~\ref{subsec:metricComparaison}).
		While better images were obtained with more $j_i^{\textrm{ref}}$, the difference between 20 $j_i^{\textrm{ref}}$ and 30 $j_i^{\textrm{ref}}$ seemed negligible.
		For this study, 30 projections of reference ($j_i^{\textrm{ref}}$) were used for each pixel such that they uniformly sampled their respective projection space.

	\subsection{ Comparison of the models }
	\label{subsec:metricComparaison}
		The images obtained with the TOF-PET log-likelihood MLEM reconstruction (\textit{MLEM}), the direct backprojection of the data with TOF Gaussian kernel (\textit{backprojection}) and two versions of the PADE model were compared in term of contrast and noise.
		The first PADE version (\textit{PADE}$_{\textrm{opt}}$) was defined with the weights $(\gamma_1, \gamma_2)$ that offered the best performance given the choice of metrics for this study. 
		The second PADE version was defined with $\gamma_1 = 0.0$ (i.e. no UDE penalization) and the minimum value of $\gamma_2$ ensuring a divergence in the number of expected coincidences lower than 1\%.
		This version will be referred to as \textit{Extended} since it is the TOF-PET log-likelihood model extended to the new parameterization.
	
		The four models were compared using the contrast recovery coefficients (CRC) and the coefficients of variation (COV) for the five smallest spots and the background, i.e. body of the phantom excluding the spots.
		The true mean value of the spots and the background was extracted from the groundtruth of each simulation.
		The pixels associated to a given region of interest were extracted using the groundtruth, with pixels subjected to significant partial volume effect excluded from the evaluation of the metrics. 
		The CRC ratio was computed as $\frac{\mu_{\textrm{spots}}}{\mu_{\textrm{bkg}}} / \textrm{CRC}^{\textrm{true}}$ where $\mu_{\textrm{spots}}$ was the mean value of the pixels extracted from the spots of interest, $\mu_{\textrm{bkg}}$ was the mean value of the pixels extracted from the background (i.e., body of the phantom excluding the spots) and $\textrm{CRC}^{\textrm{true}}$ was the CRC obtained from the simulation groundtruth over the same spots of interest.
		The background recovery ratio was defined as $\mu_{\textrm{bkg}} / \mu_{\textrm{bkg}}^{\textrm{true}}$.
		The COV was defined as $\frac{\sigma_{\textrm{ROI}}}{\mu_{\textrm{ROI}}}$ with $\sigma_{\textrm{ROI}}$ being the standard deviation of the pixels extracted from a region of interest (the spots of a given size or the background).
		The evolution of the metrics over iterations was analyzed since the convergence rate of the four models might differ.
		
		The weights $(\gamma_1, \gamma_2)$ for the \textit{PADE}$_{\textrm{opt}}$ model were obtained with a grid search, enabled by using the GNU parallel software~\cite{Tange2011a}.
		Values of $10^{-3}$ to $10^{5}$ for $\gamma_1$ and values of $10^{-5}$ to $10^{2}$ for $\gamma_2$ with multiplicative steps of ten were considered.
		For each combination of $(\gamma_1, \gamma_2)$, the solutions obtained over 250 iterations of the solver were registered for the ten simulations. 
		The mean CRC and COV values were extracted and the $(\gamma_1, \gamma_2)$ pair yielding the highest CRC ratio for the smallest spots was selected.
		$(\gamma_1, \gamma_2)$ pair were excluded if the CRC ratio of any group of spots were overestimated by more than 10\%, if the mean bias over the background was larger than 0.5 or if the number of expected coincidences deviated by more than 1\%.
		The optimum $(\gamma_1, \gamma_2)$ pair for the \textit{PADE}$_{\textrm{opt}}$ model were found to be (100.0, 10.0).
		As for the \textit{Extended} model, the optimal $\gamma_2$ was found to be 0.01.

\section{Results}
\label{sec:results}
\begin{figure}
		\centering
		\subfloat[Dataset groundtruth]{%
			\label{fig:im_gt}\includegraphics[width=.44\linewidth]{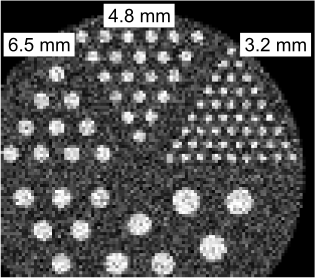}}\hfil
		\subfloat[\textit{Backprojection}]{%
			\label{fig:im_BP}\includegraphics[width=.44\linewidth]{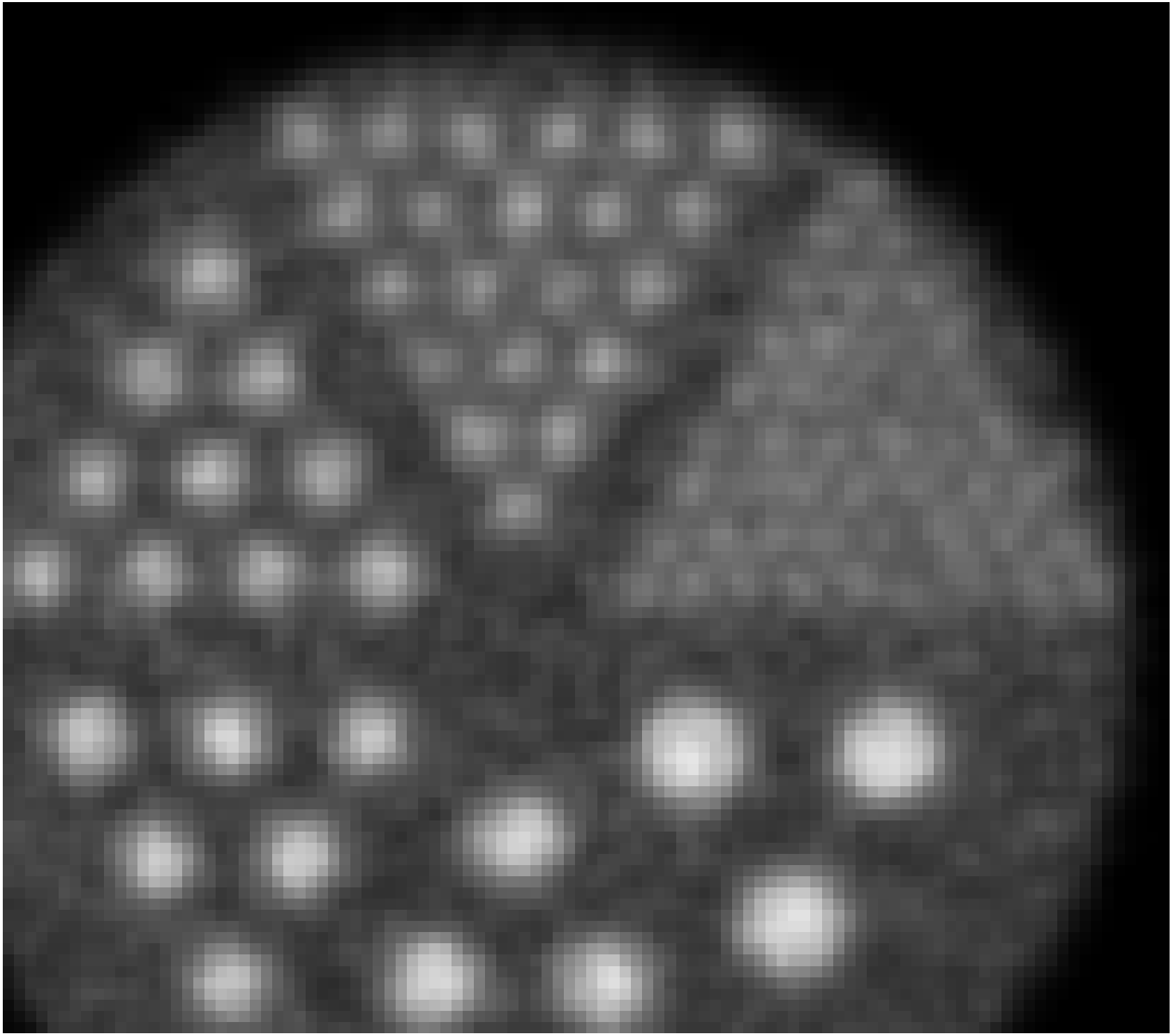}}\\
		\subfloat[\textit{MLEM}, 6 iterations]{%
			\label{fig:im_mlem_06}\includegraphics[width=.44\linewidth]{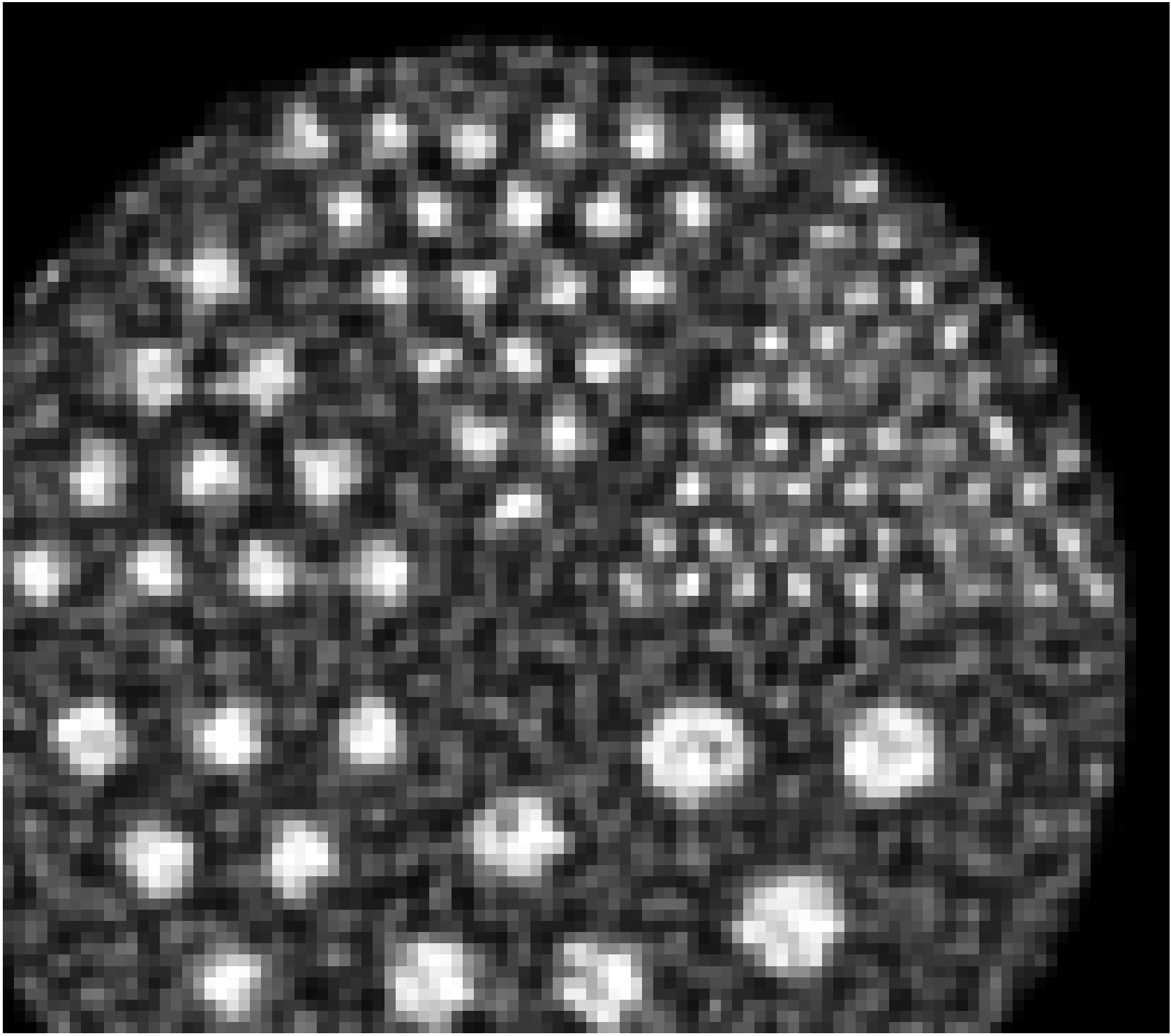}}\hfil
		\subfloat[\textit{MLEM}, 10 iterations]{%
			\label{fig:im_mlem_10}\includegraphics[width=.44\linewidth]{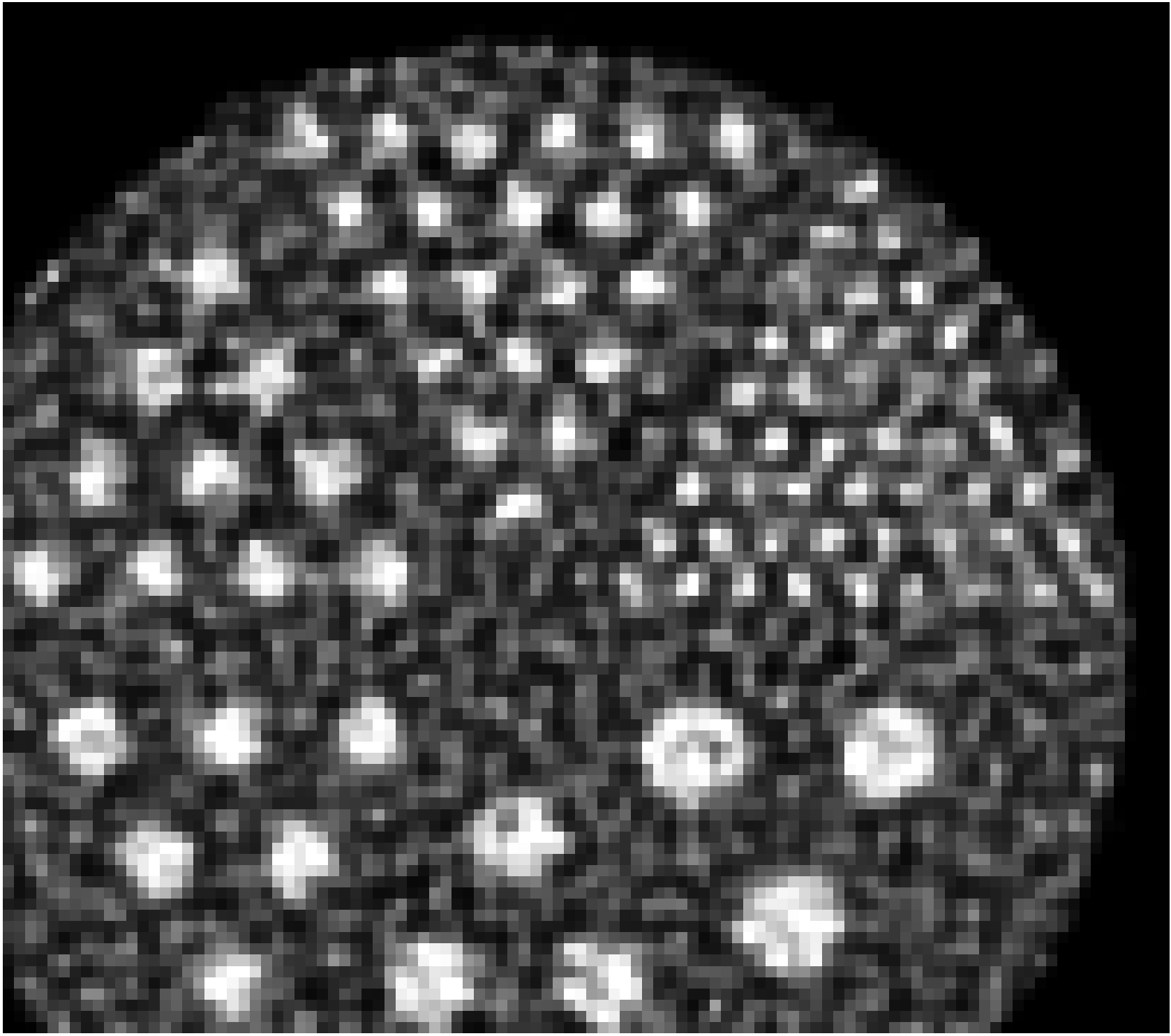}}
		\caption[Visual intuition of the effect of low counts in TOF-PET reconstructions]{%
		 	Cropped version of four images built from one of the simulations, shown with the same linear gray scale: \protect\subref{fig:im_gt} true distribution of the coincidences events; \protect\subref{fig:im_BP} image reconstructed from the backprojection of the data using the TOF Gaussian kernel; \protect\subref{fig:im_mlem_06} and \protect\subref{fig:im_mlem_10} images obtained after 6 and 10 iterations, respectively, with the MLEM algorithm.
		}%
		\label{fig:im_recon_mlemIntuition}
	\end{figure}

	In Fig.~\ref{fig:im_recon_mlemIntuition}, zoomed versions of four images obtained from one of the simulations are shown to provide some insight of the performance provided by classical approaches.
	One of the characteristics of low-count reconstruction is the inherent statistical variations inside regions of interest, as can be observed from the groundtruth (Fig.~\ref{fig:im_gt}).
	Fig.~\ref{fig:im_BP} shows the image obtained from the \textit{Backprojection} model.
	Due to the excellent TOF resolution, most of the spots were resolved albeit with a low contrast, especially for the smallest spots.
	The spots were better resolved in the image obtained with 6 MLEM iterations (Fig.~\ref{fig:im_mlem_06}) but correlated statistical noise in the background, not present in the groundtruth, was observed.
	This is an example of the images that were used to initialize the \textit{PADE}$_{\textrm{opt}}$ model, where their spots resolvability allows an efficient initialization of $\omega$ (see~\eqref{eq:omegaWeight}).
	With 10 MLEM iterations (Fig.~\ref{fig:im_mlem_10}) the spots seemed to have a better contrast but the statistical noise is further amplified, both in the spots and the background.
	These observations for the MLEM images are in agreement with the known interplay between the convergence acceleration induced by better TOF, the effect of object size on contrast recovery rate and the noisy nature of the optimal solution for the PET log-likelihood model.
	These images illustrate that even with excellent TOF resolution, low-count reconstruction remains a challenge that impedes image quality.
	
\begin{figure}
		\centering
		\includegraphics[width=\columnwidth]{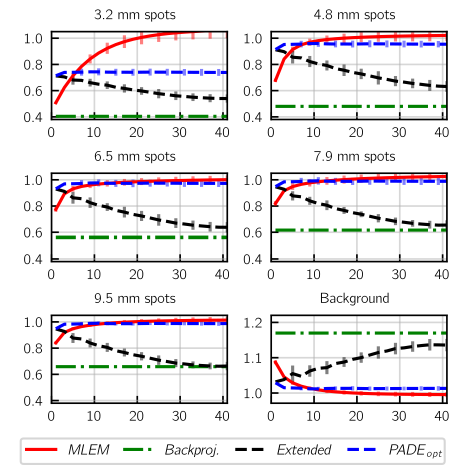}
		\caption[The CRC ratio of the five smallest spots and the background recovery ratio as a function of the number of iterations for the four models]{%
			The CRC ratio of the five smallest spots and the background recovery ratio (lower right) as a function of the number of iterations are shown for the four models.
			The values for the \textit{Backprojection} model are displayed as horizontal lines since the algorithm is non-iterative.
			Error bars ($\pm2\sigma$) show the variability over the ten repetitions.
			The error bars for the \textit{Backprojection} model are too small to be visible.
		}%
		\label{fig:crc}	
	\end{figure}

	Fig.~\ref{fig:crc} compares the CRC ratios of the five smallest spots and the background recovery ratio for the four models.
	Only the first 40 iterations out of 250 are shown since most of them converge towards stable values afterward.
	Overall, the \textit{MLEM} model seems to outperform the other models.
	Optimal values of CRC ratios and background recovery ratio are reached, except for the smallest spot for which an overestimation is observed after 20 iterations.
	The worst variations over the 10 repetitions were observed with the \textit{MLEM} model for the two smallest spots. 
	Furthermore, the image obtained at 10 MLEM iterations, see Fig.~\ref{fig:im_mlem_10}, had already been much affected by the noise, making the progress in CRC ratio afterward less conclusive.
	The \textit{Backprojection} model had the worst CRC ratios and background recovery ratio, as expected from the blurred image in Fig.~\ref{fig:im_BP}.
	However, it was also the most stable across repetitions which was expected with the choice of backprojection kernel.
	The \textit{Extended} and \textit{PADE}$_{\textrm{opt}}$ models both had a good start since they were initialized with the 6$^{\textrm{th}}$ iteration of the MLEM reconstruction.
	For the \textit{Extended} model, the metrics in Fig.~\ref{fig:crc} worsened with iteration, stabilizing at values a little better than those of the \textit{Backprojection} model. 
	The \textit{PADE}$_{\textrm{opt}}$ model achieved some gain in CRC ratio and background recovery ratio for the first few iterations and remained stable afterward.
	While it converged very rapidly ($<6$ iterations) to stable values, it did not reach the optimal values for the two smallest spots, especially with the smallest one where it did not reach 80\% of the true contrast recovery.
	
\begin{figure}
		\centering
		\includegraphics[width=\columnwidth]{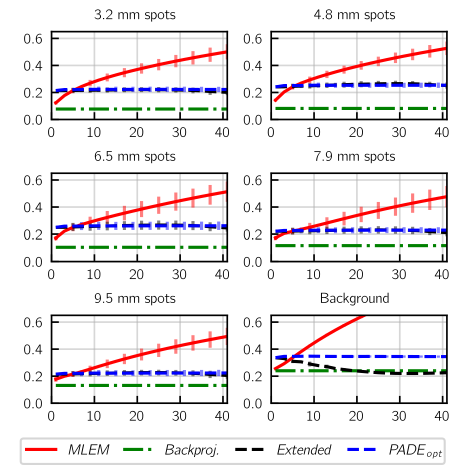}
		\caption[The COV of the five smallest spots and the background as a function of the number of iterations for the four models]{%
			The COV of the five smallest spots and the background as a function of the number of iterations are shown for the four models.
			Error bars ($\pm2 \sigma$) show the variability over the ten repetitions.
			The error bars for the \textit{Backprojection} model are too small to be visible.
		}%
		\label{fig:cov}	
	\end{figure}

	Fig.~\ref{fig:cov} compares the COV of the five smallest spots and the background for the four models.
	Only the first 40 iterations of 250 are shown since most of them are stable afterward except for the \textit{MLEM} model where it continued to increase.
	The best model for this metric was the \textit{Backprojection} model, most likely due to the smoothing effect of the TOF Gaussian kernel.
	The trends of the \textit{Extended} model differed from what was observed with the CRC metric, here remaining mostly equivalent to the \textit{PADE}$_{\textrm{opt}}$ model.
	The \textit{PADE}$_{\textrm{opt}}$ model remained stable with iterations, suggesting that the gain in CRC was achieved without increasing much the noise. 
	The \textit{MLEM} model was notably more unstable than the other models across repetitions.
	At the 10$^{\textrm{th}}$ iteration, the image produced by the \textit{MLEM} model was noisier than the one of the \textit{PADE}$_{\textrm{opt}}$ model, especially in the background.
	
\begin{figure}
		\centering
		\subfloat[Dataset groundtruth]{%
			\label{fig:im_gt_p2}\includegraphics[width=.44\linewidth]{images/groundTruthData_annoted.png}}\hfil
		\subfloat[\textit{Extended}, 40 iterations]{%
			\label{fig:im_pade_ext}\includegraphics[width=.44\linewidth]{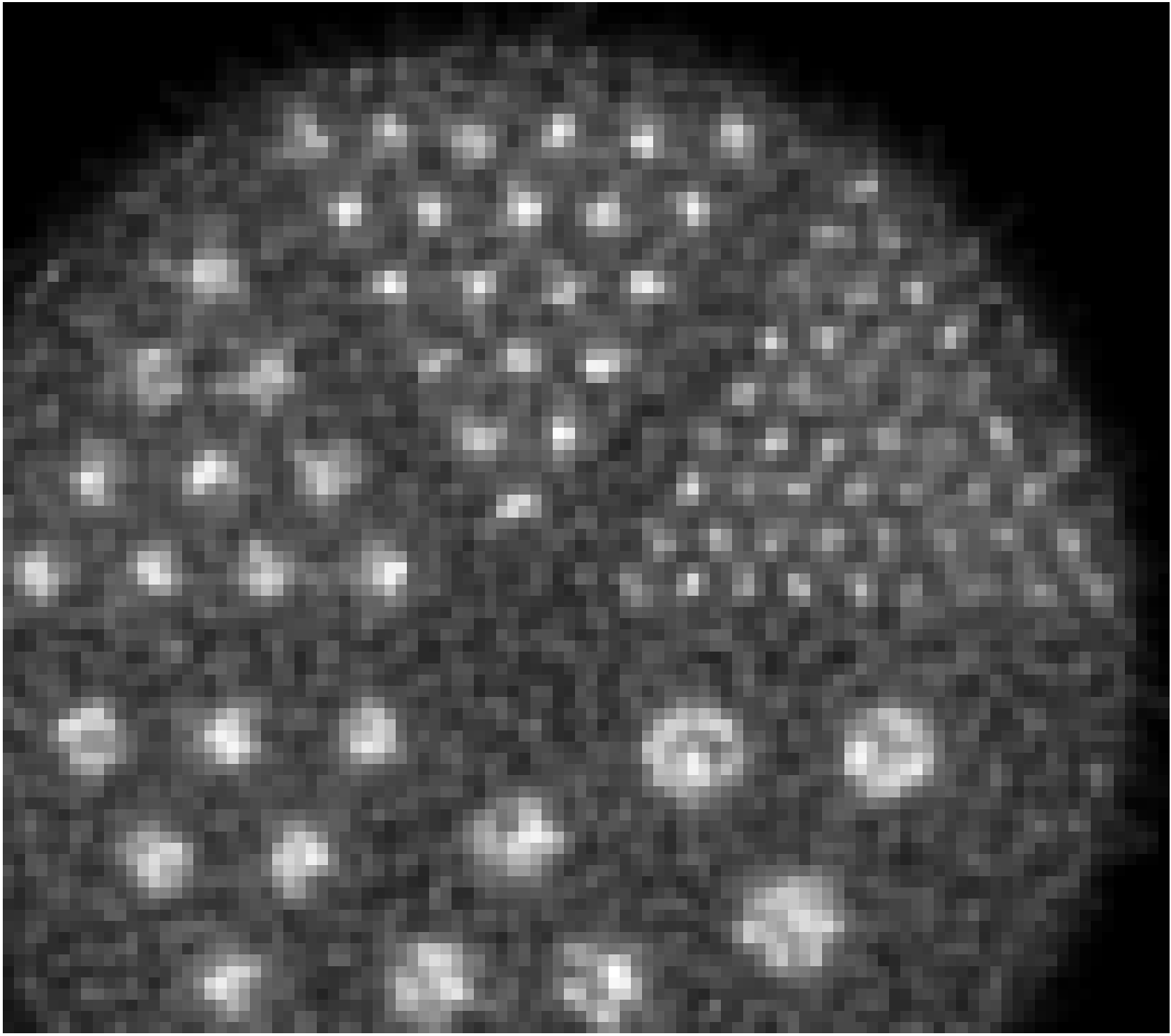}}\\
		\subfloat[\textit{PADE}$_{\textrm{opt}}$, 40 iterations]{%
			\label{fig:im_pade_best}\includegraphics[width=.44\linewidth]{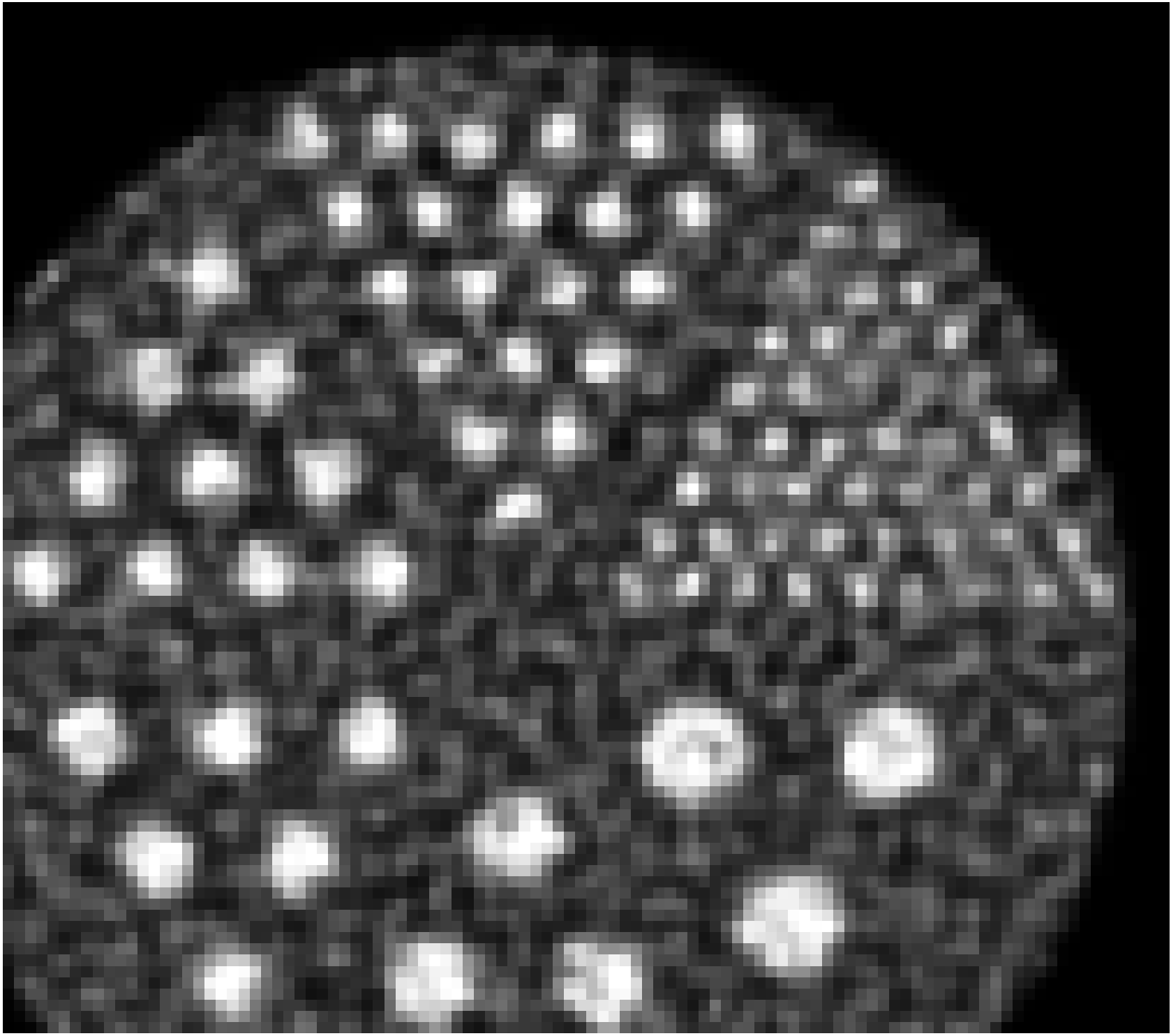}}\hfil
		\subfloat[\textit{MLEM}, 10 iterations]{%
			\label{fig:im_mlem_10_p2}\includegraphics[width=.44\linewidth]{images/MLEM_crcMoins_10.png}}
		\caption{%
			Cropped version of four images built from one of the simulations, shown with the same linear gray scale: \protect\subref{fig:im_gt_p2} true distribution of the coincidences events; \protect\subref{fig:im_pade_ext} image reconstructed with the \textit{Extended} model; \protect\subref{fig:im_pade_best} image reconstructed with the \textit{PADE}$_{\textrm{opt}}$ model and \protect\subref{fig:im_mlem_10} image reconstructed with the MLEM algorithm.
		}%
		\label{fig:im_recon_comp}
	\end{figure}

	In Fig.~\ref{fig:im_recon_comp}, a zoomed version of four images obtained from one of the simulations are shown.
	It consists of the simulation groundtruth compared to images obtained with the \textit{PADE}$_{\textrm{opt}}$, \textit{Extended} and \textit{MLEM} models.
	We chose the 40$^{\textrm{th}}$ iteration for the \textit{Extended} model (Fig.~\ref{fig:im_pade_ext}) since it was where the CRC ratios stabilized, and the 40$^{\textrm{th}}$ iteration of the \textit{PADE}$_{\textrm{opt}}$ model (Fig.~\ref{fig:im_pade_best}) to emphasize that, even with far more iterations than the \textit{MLEM}, the \textit{PADE}$_{\textrm{opt}}$ model did not produce a highly noisy solution.
	The 10$^{\textrm{th}}$ iteration of the MLEM was chosen since it was where most of the spots CRC ratios were around 1.0.
	The \textit{Extended} image has lower contrast and resolving power than the other models.
	This observation indicates that even with excellent TOF resolution, neglecting the UDE property has a detrimental effect on image contrast.
	The noise texture in the background of the \textit{PADE}$_{\textrm{opt}}$ and \textit{MLEM} images, Fig.~\ref{fig:im_pade_best} and Fig.~\ref{fig:im_mlem_10_p2} respectively, was similar except in intensity.
	This highlights that the initialization of \textit{PADE}$_{\textrm{opt}}$, shown in Fig.~\ref{fig:im_mlem_06}, had a strong impact on the solution of the proposed model. 
	The \textit{PADE}$_{\textrm{opt}}$ image appears less noisy than the \textit{MLEM} image, which seems to enhance its resolving power of the spots.
	Even with far more iterations, the \textit{PADE}$_{\textrm{opt}}$ image remained less noisy than the \textit{MLEM} image.
	
\begin{figure}
		\centering
		\subfloat[Polar distributions of $\phi_{:,i}^{40}$ (cyan and purple) and $\phi_{:,i}^*$ (red and green) for two choices of $j_i^{\textrm{ref}}$]{%
			\label{fig:udeAnalysis_polar}\includegraphics{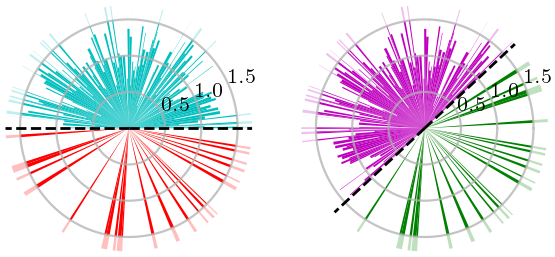}}\\
		\subfloat[Value of $\mathcal{W}_i(\phi, j_i^{\textrm{ref}})$ for 30 $j_i^{\textrm{ref}}$ for $\phi_{:,i}^{40}$ and $\phi_{:,i}^*$]{%
			\label{fig:udeAnalysis_values}\includegraphics{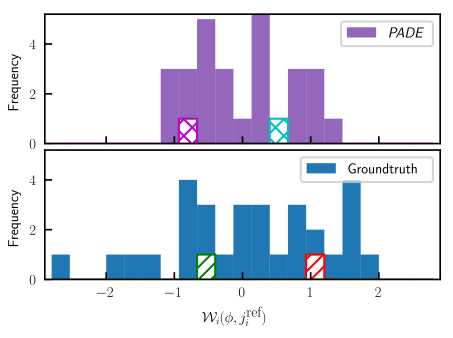}}
		\caption[Comparison of the projection-wise distribution of the proposed implementation of the PADE approach and the groundtruth of the simulation]{%
			Comparison of $\phi_{:,i}^{40}$ and $\phi_{:,i}^*$ for one of the datasets.
			$\phi^{40}$ is the 40$^{\textrm{th}}$ iteration of the \textit{PADE}$_{\textrm{opt}}$, $\phi^*$ is the dataset groundtruth and $i$ is the index of a pixel inside a 9.5~mm spot.
			$\phi_{:,i}^*$ is composed of ones and zeros and $\sum_j \phi_{j,i}^* = 24$.
			\protect\subref{fig:udeAnalysis_polar} shows their polar distributions, relative to their expected value if they followed exactly the UDE property, for two choices of $j_i^{\textrm{ref}}$, shown as a dashed black line.
			The $\tilde{\jmath}^{\textrm{th}}$ bar is equal to 1.0 if $\phi_{\tilde{\jmath},i} = (\sum_j \phi_{j,i}) R_{\tilde{\jmath},i}$.
			The cyan and purple distributions are from $\phi_{:,i}^{40}$ while the red and green ones are from $\phi_{:,i}^*$.
			For easier comparison, $\phi_{:,i}^{40}$ and $\phi_{:,i}^*$ are shown in the same graph.
			The parts that exceed 1.5 were shaded to indicate that their full height might not be visible.
			\protect\subref{fig:udeAnalysis_values} shows the values of $\mathcal{W}_i(\phi, j_i^{\textrm{ref}})$ for the 30 $j_i^{\textrm{ref}}$ when applied to $\phi_{:,i}^{40}$ (top) and $\phi_{:,i}^*$ (bottom).
			The values of $\mathcal{W}_i(\phi, j_i^{\textrm{ref}})$ for the two $j_i^{\textrm{ref}}$ illustrated in \protect\subref{fig:udeAnalysis_polar} were added with their respective color.
			The hatching pattern x was used for $\phi_{:,i}^{40}$ while the slash pattern was used for $\phi_{:,i}^*$.
		}%
		\label{fig:udeAnalysis}	
	\end{figure}

	The projection-wise distribution of the counts inside one pixel is compared in Fig.~\ref{fig:udeAnalysis} for the image obtained with 40 iterations of the \textit{PADE}$_{\textrm{opt}}$ model ($\phi_{:,i}^{40}$) and the groundtruth ($\phi_{:,i}^{*}$) for one of the datasets.
	The compared pixel, $i$, was inside one of the 9.5~mm spots and it had 358 valid projections.
	$\phi_{:,i}^{*}$ consisted of zeros and ones, for a total of 24 counts.
	Fig.~\ref{fig:udeAnalysis_polar} shows the polar distributions of $\phi_{:,i}^{*}$ and $\phi_{:,i}^{40}$, relative to their expected value if they exactly followed the UDE property, for two choices of $j_i^{\textrm{ref}}$.
	The theoretical momentum, i.e., the momentum of a distribution that follows exactly the UDE property, would split these four polar distributions exactly in half.
	$\phi_{:,i}^{*}$ seems nearly evenly distributed on both sides of the theoretical momentum for both $j_i^{\textrm{ref}}$ (see the red and green polar distributions). 
	The same observation applies to $\phi_{:,i}^{40}$ (see the cyan and purple polar distributions).
	In other words, the momentum of $\phi_{:,i}^{*}$ and $\phi_{:,i}^{40}$ should be close to the theoretical momentum for those two $j_i^{\textrm{ref}}$, suggesting they follow a uniform distribution. 
	$\phi_{:,i}^{40}$ was clearly not sparse, contrary to $\phi_{:,i}^{*}$, and we note that its values fluctuate around 1.0, which means that $\phi_{j,i}^{40} \approx (\sum_j \phi_{j,i}^{40}) R_{j,i}$.
	This last observation suggests that the classical interpretation of the UDE property was strongly enforced with the \textit{PADE}$_{\textrm{opt}}$ model.
	Fig.~\ref{fig:udeAnalysis_values} shows the histograms of the values of $\mathcal{W}_i(\phi_{:,i}^{40}, j_i^{\textrm{ref}})$ and $\mathcal{W}_i(\phi_{:,i}^{*}, j_i^{\textrm{ref}})$ over the 30 sampled reference projections ($j_i^{\textrm{ref}}$).
	Note that values outside of the $[-0.5, 0.5]$ range are possible since the momentum was defined as a weighted sum and not a mean.
	The values of $\mathcal{W}_i()$ for the distributions shown in Fig.~\ref{fig:udeAnalysis_polar} were highlighted to provide a visual assessment of the metric. 
	For example, $\mathcal{W}_i()$ indicated that the central tendency of the four polar distributions were different in amplitude and sign (the $x$ position of the bin with a hatched pattern indicates the value of  $\mathcal{W}_i()$ for the corresponding polar distribution color).
	While the two histograms are mostly within the same range, the values associated to $\phi_{:,i}^{*}$ (blue) were more dispersed than those of $\phi_{:,i}^{40}$ (violet).
	Thus, $\mathcal{W}_i()$ shows that $\phi_{:, i}^{40}$ better follows the UDE property than $\phi_{:, i}^{*}$, which again suggests that the \textit{PADE}$_{\textrm{opt}}$ model strongly enforces the UDE property.

\section{Discussion}
\label{sec:disc}
	The goal of this study was to demonstrate the potential of the PADE paradigm, enforcing the uniform distribution of emission (UDE) property, for the reconstruction of low counts ultra-fast TOF-PET acquisitions.
	The proposed implementation of the PADE approach, \textit{PADE}$_{\textrm{opt}}$, achieved similar CRC ratios as the \textit{MLEM} model in most regions of interest with better noise property and its solutions were more stable.
	However, we can not conclude that it outperforms the classical log-likelihood PET reconstruction model, especially when looking at the CRC ratios obtained on the smallest spots.
	A closer look taken on the solution obtained with the \textit{PADE}$_{\textrm{opt}}$ model showed that the proposed model still enforces strongly the classical interpretation of the UDE property, i.e. $\phi_{j,i} \approx R_{j,i} \lambda_i$.
	Also, the performance of the \textit{Extended} model showed that removing completely the UDE property from the reconstruction scheme is disadvantageous, even if ultra-fast TOF is available. 
	These results suggest that the PADE approach is promising for ultra-fast TOF reconstruction of low-count acquisitions, but further investigation will be needed to ascertain its full potential, especially regarding the UDE penalization term.
	
	\subsection{Review of the \textit{PADE}$_{\textrm{opt}}$ model}
	\label{subsec:reviewImplementation}
		The MLEM algorithm, which implicitly enforces non-negativity and expected counts stability, provides quite good results in most cases, if stopped judiciously.
		Given its ease of use, it is understandable that other approaches for PET reconstruction often remain mostly academic. 
		The requirement of finding the appropriate penalization weights $\gamma_1$ and $\gamma_2$, which are most likely data-dependent, make the \textit{PADE}$_{\textrm{opt}}$ model less straightforward than the classical log-likelihood model.
		Since $\gamma_2$ is used to preserve the expected number of counts, its calibration can be automated by searching the smallest value that enforces this property.
		However, the weight factor enforcing UDE, $\gamma_1$, is likely to pose a challenge, as it is the case for other PET algorithms that include a regularization scheme~\cite{qi2006iterative}.
		Further investigation is required to understand how critical is the calibration of $\gamma_1$ and its dependence to the data. 
		The solution proposed for the maximum \textit{a posteriori} model in~\cite{reader2020bootstrap}, where an automated calibration was proposed, might be applicable for the \textit{PADE}$_{\textrm{opt}}$ model.
		
		The TOF-less log-likelihood model was underdetermined in this study (8,544 projections vs. 16,384 pixels).
		While our choice of TOF binning does increase the system sampling well beyond the number of pixels, it does not ensure that the resulting system is overdetermined.
		We expect that the \textit{PADE}$_{\textrm{opt}}$ model would perform better with a scanner having more projections.
		Indeed, the freedom provided by relaxing the UDE property should shine when the number of valid projections is far larger than the number of counts in a voxel. 
		However, the computational burden increases with the number of projections, which has limited our capability to calibrate the \textit{PADE}$_{\textrm{opt}}$ model on scanners with more detectors.
		For both the \textit{PADE}$_{\textrm{opt}}$ model and the \textit{MLEM} model, the 3.2~mm spots were mostly resolved even with detectors being 8~mm in width.
		This was shown possible with the MLEM algorithm~\cite{toussaint2020improvement} and it is reassuring that the PADE approach seems to share the same feature.
		Yet, these small structures combined with a low-count acquisition have posed a challenge for the \textit{MLEM} model.
		This is highlighted by its CRC ratio, where a value of 1.0 is reached briefly before being overestimated, well after the noise started to significantly spoil the image.
		Thus, comparison of the models using the CRC of the smallest spot is less straightforward and the \textit{MLEM} model might not be an adequate baseline in that case.
	
		The increase in degrees of freedom provided by the PADE paradigm combined with the interplay between the three terms in~\eqref{eq:padeModel} also means that the hyperparameters have a significant impact on the optimal solution of the model, which can deviate significantly from the groundtruth.
		Thus, an automated study of several hyperparameters based on one quality metric can be misleading and a reduction of the number of unknowns was required. 
		We simplified the study by defining a phantom with only two level of concentrations: the background with a low value and the spots with a high value.
		Thus, the threshold in~\eqref{eq:omegaWeight} only needed to be able to distinguish these two activity levels.
		As for the value of the $\omega_i$, they were mostly applied to pixels inside the spots, due to the threshold and the initialization. 
		Their impact was therefore limited to a subset of pixels that should have similar number of counts. 
		Even if the optimal $\omega$ was not directly the number of counts per pixels, this was compensated by $\gamma_1$ since it was calibrated to achieve the best performance.
		
		We had observed that having $\omega_i \approx \sum_j \phi^*_{j,i}$, with $\phi^*$ being the groundtruth, provided encouraging results.
		However, $\phi^*$ being unknown in practice, we opted to update $\omega$ with the previous iteration (see~\eqref{eq:omegaWeight}) and use a low iteration MLEM to initialize the reconstruction (see~\eqref{eq:initialization}).
		While this approach makes it possible for $\omega$ to reach the desired value, it also means that the objective function is modified at each iteration.
		Considering that the L-BFGS-B algorithm employs past iterations to create an approximation of the Hessian matrix of the objective function, it is possible that its convergence properties are weakened or lost. 
		Nevertheless, the \textit{PADE}$_{\textrm{opt}}$ model was initialized with a pretty good first guess, which might limit the impact of modifying $\omega$ on the solver.
		The stability of the CRC ratios and COV over iterations for the \textit{PADE}$_{\textrm{opt}}$ model goes on longer than the ten iterations that the L-BFGS-B algorithm uses for the Hessian approximation.
		This suggests that the solution obtained with the current iterative scheme is optimal for the \textit{PADE}$_{\textrm{opt}}$ model.
		Moreover, since only a low iteration MLEM is needed with ultra-fast TOF to discern most structures, it might be possible to design an automatic criterion that is independent of the groundtruth.
		Further investigations will be required to evaluate the impact of initialization on the performance of the \textit{PADE}$_{\textrm{opt}}$ model. 
		Lastly, another choice of general solver might provide a better behavior.
		For example, the ADMM algorithm breaks the optimization problem into smaller ones that are easier to solve and it might be more efficient at dealing with a model defined as three terms competing with each other~\cite{teng2016admm,6825888}.

	\subsection{Potential of the PADE paradigm}
	\label{subsec:futurePade}	
		The main novelty of the proposed model is in the interpretation the UDE property which, in the classical log-likelihood model, is enforced as $\phi_{j,i} = R_{j,i} \lambda_i, \forall j,i$ with the system matrix.
		The goal of $\mathcal{V}_i()$ is to promote solutions that follow the UDE property without requiring each $\phi_{j,i}$ to be close to its theoretical expected value ($R_{j,i} \lambda_i$).
		This is of particular interest when the number of counts per pixel is low.
		The implementation of $\mathcal{V}_i()$ in~\eqref{eq:geoPenalMetric} uses the fact that the momentum of an infinitely sampled circular uniform distribution should be at the center of its domain for any choice of $j_i^{\textrm{ref}}$.
		The momentum is a central tendency metric enabling the characterization of a pixel emission distribution as a whole, without relying on the individual values of $\phi_{j,i}$.
		However, the proposed implementation of $\mathcal{V}_i()$ is cumbersome in terms of computational resources even with 30 $j_i^{\textrm{ref}}$.
		Further investigation will be needed to simplify or to improve $\mathcal{V}_i()$ in terms of flexibility over the weights.
		The threshold applied on $\omega$ in~\eqref{eq:omegaWeight} is used to ensure that $\mathcal{V}_i()$ does not overshadow the log-likelihood term for pixels with only a few counts.
		It remains to be shown that this implementation of $\mathcal{V}_i()$ is efficient for the PADE paradigm.
		
		The analysis of Fig.~\ref{fig:udeAnalysis} seems to indicate that the proposed implementation of the PADE approach enforced something akin to the classical interpretation of the UDE property ($\phi_{j,i} = R_{j,i} \lambda_i$).
		While this could be due to a lack of flexibility from $\mathcal{V}_i()$ (that is, it still mainly promotes the classical interpretation of the UDE property), there is another probable explanation.
		The proposed model is initiated using~\eqref{eq:initialization} which is based on the classical interpretation of the UDE property.
		Thus, it is possible that the observations made in Fig.~\ref{fig:udeAnalysis} are caused by the reconstruction model being underdetermined and that the initialization projection-wise pattern, i.e. $\phi_{:, i}^0$ remains, in parts, similar over the iterations.
		As claimed in section~\ref{subsec:reviewImplementation}, the camera configuration studied here is clearly underdetermined in its TOF-less form, which could be handled by having a camera configuration with more detectors.
		However, the size of $\phi$ would also grow with the number of detectors, albeit less quickly since the number of deactivated parameters (see~\eqref{eq:constraints}) would also increases.
		In other words, it is possible that, even with ultra-fast TOF, the parametrization of $\phi$ provides too many degrees of freedom for the data fit term, i.e., the log-likelihood, to promote solutions that are akin to the groundtruth. 
		Thus, a better correspondence to the groundtruth could be expected if sparse solutions were enforced, which might be possible with ultra-fast TOF.
		The simplest approach would consist of using a voxel-wise sparsity penalty term, but the calibration would then be an even more complex interplay between the strength of each term. 
		Algorithmic approaches to enforce sparsity, such as FISTA~\cite{beck2009fast}, might be preferable, if it can be adapted to the PADE model.
		Integer optimization is another alternative that might benefit the proposed model.
		The parametrization of the PADE model enables a more accurate use of the integer nature of counts by circumventing the $R_{ji} \lambda_i$ approximation in the log-likelihood term.
		While the number of variables for the PADE model is definitely in the high end for integer programming, multiple heuristic methods have been successfully developed for diverse integer optimization problems and an in-depth investigation of its state-of-the-art might provide fruitful solutions~\cite{kumar2010fifty,glover1997general}.
		Other approaches could also be investigated, such as origin ensembles that was already applied to TOF-PET reconstruction~\cite{wulker2015time} or machine learning methods which have also shown their potential for PET reconstruction~\cite{gong2019machine}.
		
		We expect that the PADE model would perform even better for 3D reconstructions.
		The histogram being even sparser in 3D, more valid projections could be removed for each voxel.
		Thus, the resulting model should be easier to solve for 3D reconstructions.
		In this study, the mean percentage of valid projections that can be removed for pixels inside the phantom was around 50\%, even though the number of detectors was low (320 for one ring of a clinical size scanner). 
		If we double the number of detectors, the mean percentage rises to 80\%, which results in 10\% less valid projections being kept than the configuration studied (50\% of $\approx$320 vs 80\% of $\approx$640).

\section{Conclusion}
	The foundations of a new approach, Parameterizing the Angular Distribution of Emission (PADE), have been developed to deal with the difficulties associated with low counts in PET reconstruction, particularly in the context of ultra-fast TOF resolution.
	The PADE approach offers greater degrees of freedom than the classical log-likelihood model by relaxing the interpretation of the uniform distribution of emission, usually encoded in the system matrix.
	This increases the number of variables significantly, which cannot be exploited fully in a classical PET setting. 
	Ultra-fast TOF reconstruction with low counts represents a best case scenario for the PADE approach since the number of variables can be reduced by the excellent TOF resolution without loss in degrees of freedom. 
	The model implemented in this paper provides some gains in image quality, albeit limited, over the log-likelihood model, which constitutes a proof of concept of the PADE approach and thus suggests that this new approach has potential benefits warranting further investigations.
	The core idea of the PADE approach is promising and we believe that an efficient and stable implementation of this idea can be achieved.

\section*{Acknowledgments}
	The authors gratefully acknowledge Jean-Baptiste Michaud for providing access to his allocated computational cluster ressources.


\bibliographystyle{IEEEtran}

\bibliography{main.bib}

\end{document}